# Self-amplifying Hawking radiation and its background: a numerical study

Jeff Steinhauer and Juan Ramón Muñoz de Nova

*Department of Physics, Technion—Israel Institute of Technology, Technion City, Haifa 32000, Israel*

We numerically study an analogue black hole with two horizons with similar parameters to a recent experiment. We find that the Hawking radiation exists on a background which contains a density oscillation, a zero-frequency ripple. The Hawking radiation evolves from spontaneous to self-amplifying, while the background ripple grows steadily with no qualitative change. It is seen that the self-amplifying Hawking radiation has a non-zero frequency. The background ripple appears even before the inner horizon is created, in contrast to predictions. This work is in agreement with the recent observation of self-amplifying Hawking radiation, and explains some of the features seen. In contrast to recent works, our study differentiates between the Hawking radiation observed, and the evolution of the background.

Spontaneous Hawking radiation from a black hole with one horizon should have a thermal energy distribution [1, 2]. This is also true for analogue black holes [3-14]. On the other hand, an analogue black hole with two horizons and a superluminal dispersion relation can exhibit self-amplifying Hawking radiation, or "black hole lasing" [15]. This phenomenon has been studied extensively theoretically [15-22] and experimentally [23]. The outer and inner horizons are analogous to the horizons in a charged black hole. The self-amplifying Hawking radiation appears as a growing standing wave between the horizons, oscillating with a single frequency. Due to the stochastic nature of the Hawking radiation, the standing wave is visible in the density-density correlation function computed from an ensemble of repetitions of the experiment.

It was predicted by Jain, et al. [16] that the background density between the horizons contains a "ripple", a zero-frequency wave which is a feature of the stationary background flow. The self-amplifying Hawking radiation grows upon this background. Naturally, the ripple does not appear in the density-density correlation function since it does not fluctuate.

Recently, the phenomenon of self-amplifying Hawking radiation was observed via the density-density correlation function [23]. The ripple in the background was also observed via the ensemble-averaged density profile. Inspired by the experimental observation, the Orsay group numerically studied the evolution of the self-amplifying Hawking radiation in the presence of time-dependent flow [24]. They found that due to nonlinear backaction, the self-amplifying Hawking radiation can create a zero-frequency background ripple at late times. The Trento and Maryland groups essentially studied the background ripple only [25, 26]. However, it was not clear that it was the background being studied as opposed to the black hole lasing itself. Here, we show that the self-amplifying Hawking radiation and the evolution of the background ripple are two distinct effects which evolve independently. We study both the fluctuations and the background by numerical simulation. We find that the fluctuations show a transition from spontaneous to self-amplifying Hawking radiation, while the background ripple grows continually across the transition. The black hole lasing is seen to have a frequency with a non-zero real part. Furthermore, it was thought that the background ripple was associated with Bogoliubov-Čerenkov emission from the inner horizon [25, 26], but we see that the background ripple appears well before the formation of the inner horizon, at least for the parameters studied here.

Fig. 1 shows the dispersion relation in the supersonic region between the horizons. The zero-frequency mode $k_z$ is indicated. This is the mode which appears as a ripple in the background. The black hole lasing mode on the other hand, is a standing wave between the $P$ and $in_-$ modes, with finite frequency $\omega_L$, where $\omega_L$ is the real part of the complex frequency. The $P$ mode is the negative energy member of the Hawking pairs emitted from the outer black hole horizon. The $in_-$ mode is created when the $P$ mode reflects from the inner horizon. The finite cavity contains a few possible modes, and the dominant lasing mode $\omega_L$ is the mode with the lowest frequency [19]. The wavenumber of the lasing mode is $k_L = k_P - k_{in_-}$. Since $\omega_L$ is small, $k_L$ is close to $k_z$. Thus, the background contains a ripple with a wavelength which is similar to that of the self-amplifying Hawking radiation.

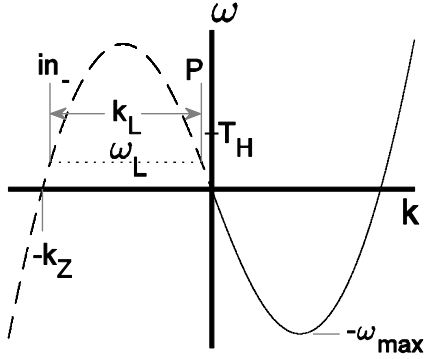

FIG. 1. The dispersion relation in the supersonic black hole laser cavity. The dashed curve is the negative norm branch. $k_z$ is the zero-frequency mode. The difference between $k_P$ and $k_{in_-}$ is the wavenumber $k_L$ of the self-amplifying Hawking radiation. $\omega_L$ indicates the frequency of the self-amplifying Hawking radiation. $T_H$ is the Hawking temperature.

It is not clear what triggers self-amplifying Hawking radiation. One proposal is that the zero-frequency mode is the trigger [25]. Here, we consider the possibility that the spontaneous Hawking radiation triggers the self-amplifying Hawking radiation. The temperature of spontaneous Hawking radiation in an analogue system is given by

$$k_B T_H = \frac{\hbar}{2\pi}\left(\frac{dc}{dx} - \frac{dv}{dx}\right) \tag{1}$$

where the flow velocity $v$ and the speed of sound $c$ determine the metric, and the derivatives are evaluated at the black hole horizon. If the spontaneous Hawking radiation is to efficiently trigger the self-amplifying Hawking radiation, then $k_B T_H$ should be larger than $\hbar\omega_L$, where $\omega_L$ is the frequency of the self-amplifying Hawking radiation.

We numerically study the self-amplifying Hawking radiation via the one-dimensional Gross-Pitaevskii equation. The parameters are similar to those of the experiment of [23]. Quantum fluctuations are added to the system soon after the formation of the outer black hole horizon. They are added by our short-pulse Bragg technique, in which a random potential is turned on for a very short time [4, 27]. The potential is filtered in $k$-space by the function $(U_k + V_k)^{-1}$, where $U_k$ and $V_k$ are the Bogoliubov amplitudes of the modes. This results in fluctuations of the correct order of magnitude both inside and outside the supersonic region. The time $t = 0$ is taken to be the time that the fluctuations are added.

Fig. 2 shows the time evolution of the black hole laser. The upper panel at each time shows the Hawking radiation fluctuations via the correlation function. The middle panel at each time shows the background density profile with its ripple. The lower panel at each time shows the flow velocity and the speed of sound. The following sequence of events is seen:

i. The outer black hole horizon forms.

ii. Spontaneous Hawking radiation appears, as well as the background ripple. The background ripple grows continuously for all later times.

iii. The inner horizon forms.

iv. Self-amplifying Hawking radiation appears.

At the 3 earliest times of Fig. 2 (the top row), spontaneous Hawking radiation is seen. It is identified by the dark band of correlations extending from the black hole horizon, indicated by "BH". Each point along this band corresponds to equal propagation times from the horizon. These are the correlations between the Hawking/partner pairs. The self-amplifying Hawking radiation becomes slightly visible in Fig. 2c and clearly visible in Fig. 2d. It is characterized by a checkerboard pattern in the correlation function. Remnants of the spontaneous Hawking radiation are seen simultaneously with the self-amplifying Hawking radiation in Fig. 2d. These remnants are located away from the black hole horizon, so they likely result from earlier emission of spontaneous Hawking radiation.

The ripple in the background is visible and steadily growing throughout the transition from spontaneous to self-amplifying Hawking radiation, as seen in the density profiles (the middle panels) of Fig. 2. This suggests that the self-amplifying Hawking radiation and the ripple in the background are separate phenomena. It has been suggested that the ripple in the background is emitted by the inner horizon as Bogoliubov-Čerenkov radiation [25]. However, the ripple in the background forms even before the inner horizon -- The ripple is seen in the middle panel of Fig. 2a, but the $v$ and $c$ curves in the lower panel do not yet cross to form an inner horizon.

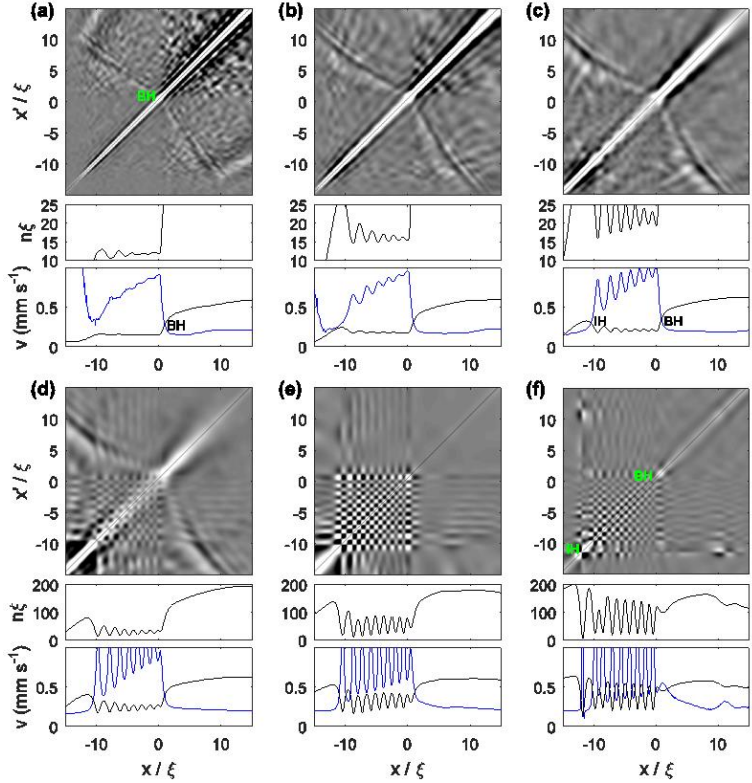

FIG. 2. Time evolution of the black hole laser. (a-f) show 40, 60, 80, 100, 140, and 200 ms after the moment when the quantum fluctuations are introduced into the simulation. The upper panel at each time is the correlation function, which shows either spontaneous or self-amplifying Hawking radiation. The grayscale in (f) corresponds to a larger range than the other panels. The middle panel at each time shows the density profile with its ripple. The lower panel at each time shows the background speed of sound $c$ (black curve) and flow velocity $v$ (blue curve).

We can study the frequency of the self-amplifying Hawking radiation by following the time evolution of each member of the ensemble separately. As seen in Fig. 1, the background ripple and the self-amplifying Hawking radiation have almost the same wavelength, so they add at a given time to form a single spatial wave. This wave contains a zero-frequency contribution from the background ripple, as well as a time-oscillating contribution from the self-amplifying Hawking radiation. At each time, the amplitude $n_{ki}(t)$ represents the magnitude of the spatial Fourier transform of the density profile in the lasing region, where $i$ indicates the member of the ensemble. Due to the zero-frequency background ripple, $n_{ki}(t)$ steadily grows, as seen in the middle panels of Fig. 2. $n_{ki}(t)$ also has a time-oscillating component due to the self-amplifying Hawking radiation. In order to discern the oscillating component, we remove the steady growth by dividing by $n_i(t)^2$, where $n_i(t)$ is the spatial average of the density in the lasing region. The $n_i(t)^2$ growth rate was predicted in [26]. Thus, we consider $\tilde{n}_{ki}(t) \equiv n_{ki}(t)/n_i(t)^2$. We then compute the power spectrum $\langle |F_i(\omega)|^2 \rangle$, where $F_i(\omega)$ is the Fourier transform of $\delta\tilde{n}_{ki}(t)$, $\delta\tilde{n}_{ki}(t) \equiv \tilde{n}_{ki}(t) - \langle\tilde{n}_{ki}(t)\rangle$, and the averages are over the ensemble.

The resulting power spectrum is shown in Fig. 3a. A shallower potential step than that of Fig. 2 is shown, resulting in a shorter lasing cavity, as for the shallowest step in Ref. [23]. A peak of non-zero frequency is clearly seen, giving the frequency $\omega_L$ of the self-amplifying Hawking radiation. Thus, the self-amplifying Hawking radiation is differentiated from the zero-frequency background ripple. Each curve in Fig. 3a is labeled by the finite slope $\xi/w$ of the potential gradient at the black hole horizon, where $w$ is the width of the potential step (the waist of the half Gaussian forming the step), $\xi = \hbar/mc_L$ is the healing length, $c_L$ is the average speed of sound in the lasing region, and $m$ is the atomic mass. Varying $\xi/w$ has two effects. Firstly, it gives control over the Hawking temperature of the black hole horizon via the hydrodynamic derivatives $dc/dx$ and $dv/dx$, by Eq. 1. These derivatives increase for increasing $\xi/w$, but they saturate for $\xi/w \gg 1$. Secondly, decreasing $\xi/w$ decreases the length of the lasing cavity. Thus, Fig. 3a shows 2 resonant modes – a higher $\omega_L$ for a longer cavity which contains 4 spatial oscillations, and a lower $\omega_L$ for a shorter cavity with 3 spatial oscillations. The lower frequency resonant peak is seen to be larger, such as in the $\xi/w = 1.5$ curve. This is not surprising since lower frequencies and shorter cavities should give faster growth of the self-amplifying Hawking radiation [19, 23]. However, for smaller slopes such as $\xi/w = 1$, the peak in Fig. 3a is seen to disappear. The cavity contains 3 spatial oscillations, but the resonance is not excited. This is consistent with the possibility that the self-amplifying Hawking radiation is stimulated by the spontaneous Hawking radiation. For the low values of $\xi/w$, the Hawking temperature may be too low to efficiently stimulate the mode with energy $\hbar\omega_L$.

The circles of Fig. 3b show the heights of the peaks in Fig. 3a as a function of $\xi/w$. As in Fig. 3a, the height of the peak increases for decreasing $\xi/w$, until the lower cutoff is reached. The amplitude squared derived from the correlation function shows similar behavior, as indicated by the squares. In contrast, the background ripple is relatively independent of $\xi/w$, as seen in the inset to Fig. 3b. Again, we see that the self-amplifying Hawking radiation and the ripple are unrelated phenomena.

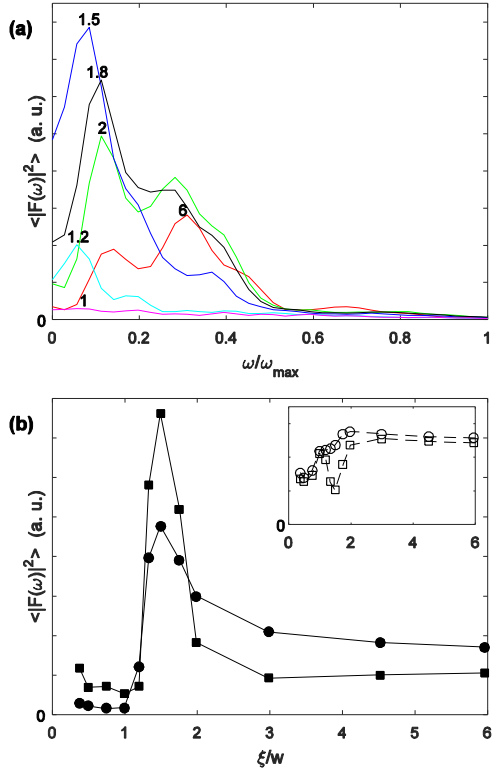

FIG. 3. The frequency of the self-amplifying Hawking radiation. (a) The resonant peaks of the self-amplifying Hawking radiation. The various curves correspond to different values of $\xi/w$. (b) The self-amplifying Hawking radiation as a function of $\xi/w$. The circles and squares indicate the amplitude of the self-amplifying Hawking radiation as derived from the time dependence and the correlation function, respectively. The relative magnitude of the two curves is arbitrary. The squares are the magnitude of the $P_2$ peak of Fig. 4h. The inset shows the amplitude squared of the background ripple. The squares indicate the amplitude squared of the average profile. The circles indicate the mean squared amplitude of the ensemble, which includes a small contribution from the self-amplifying Hawking radiation.

A recent article [25] suggested that the black hole lasing correlation function could result from the ripple in the background, in combination with technical noise. In other words, the upper panel of Fig. 2 should result from the middle panel. We can see that this picture does not apply by inspecting Fig. 2a or 2b where the spontaneous Hawking radiation of the upper panel is qualitatively different from the middle panel. Nevertheless, we can check whether this picture is at least consistent with the late time behavior, when the ripple and the fluctuations have similar spatial patterns, as in Fig. 2f. In [25], technical noise in the height of the potential step which was more than 2 orders of magnitude larger than in the actual experiment of [23] was applied. The resulting correlation function from [25] is shown in Fig. 4d. It is seen to have a crucial difference from the experimental correlation function from [23] shown in Fig. 4a. The experimental correlation function has a 2D array of white spots near the black hole horizon, within the green rectangle. Such a 2D array indicates that each member of the ensemble has the same node locations. This is a signature of a standing wave between the horizons, as in self-amplifying Hawking radiation. Fig. 4d lacks this 2D array, and thus differs from the experimental result in a crucial way. In order to see this quantitatively, Figs. 4b and 4e show the Fourier transform of the correlation patterns. Fig. 4e is missing the peak indicated by "$P_2$". This is the peak that would have resulted from the 2D array, as opposed to peak "$P_1$" which merely reflects the lines parallel to the diagonal seen in Fig. 4d. Such parallel lines indicate nodes with variable position rather than a standing wave. In other words, the $P_1$ peak is proportional to $\langle \rho_k \rho_{-k} \rangle = \langle |\rho_k|^2 \rangle$, while the $P_2$ peak has spatial phase information since it is proportional to $|\langle \rho_k \rho_k \rangle| = |\langle \rho_k^2 \rangle|$. The ratio between the $P_1$ and $P_2$ peaks in Fig. 4e is 0.14, in comparison with 0.42 in Fig. 4b. The fact that the ratio is so low in Fig. 4e shows that the ripple-plus-technical noise model does not reproduce the experimental standing wave. In contrast, the vacuum fluctuations added to the simulation in the present work do show the 2D array with a ratio of 0.76, as seen in Fig. 4h.

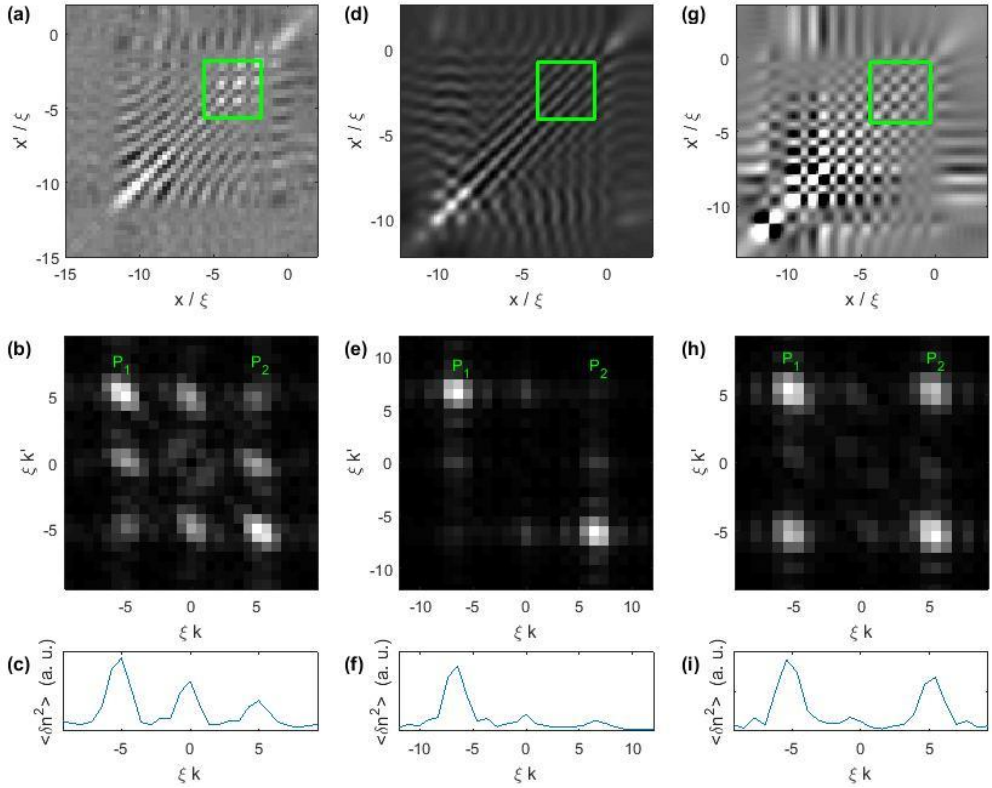

FIG. 4. The correlation pattern. (a) The experimental correlation function. The Fourier transform is computed in the green square. (b) The Fourier transform of (a). The ratio of the $P_2$ peak to the $P_1$ peak measures the stationarity of the nodes near the outer horizon. (c) The profile of the $P_1$ and $P_2$ peaks. (d-f) The simulation of [25]. (g-i) The simulation in this work.

It was suggested in Ref. [26] that this experimental scenario does not exhibit self-amplification due to the recession in time of the white hole from the black hole. This is not true for our simulations, since the self-amplifying Hawking radiation is clear in Figs. 2, 3, and 4. In order to further check the assertion, we have numerically studied the effect of the recession with constant velocity of the white hole in the theoretical black hole laser configuration described in Refs. [20, 22], where one can isolate the pure effects of self-amplification as no Bogoliubov-Čerenkov radiation is emitted from the white hole [22]. The results show that, for a wide range of white-hole velocities, the self-amplification process which gives rise to the dynamical instability of the lasing cavity still takes place. Hence, the recession of the white hole in the actual experiment should not prevent the appearance of the black-hole laser effect.

In conclusion, in order to describe the fluctuations seen in the recent experiment, one must include the non-zero frequency fluctuations associated with the black hole lasing effect. It is not sufficient to consider only the zero-frequency physics of the background ripple. We see evidence that the spontaneous Hawking radiation triggers the self-amplifying Hawking radiation. We also see that the self-amplifying Hawking radiation and the ripple in the background are separate, largely independent phenomena, despite their similar spatial patterns, as predicted in [16]. We find that the self-amplifying Hawking radiation is different from the background ripple in the following ways:

i. It (the self-amplifying Hawking radiation) appears in the density-density correlation function.
ii. It evolves from spontaneous Hawking radiation while the background ripple exhibits no qualitative change.
iii. It is seen to have a non-zero frequency.
iv. Its amplitude increases for decreasing frequency and lasing cavity length, as predicted.
v. Its amplitude seems to depend on the Hawking temperature of the outer horizon.

Both the background ripple and the self-amplifying Hawking radiation are clear in the experiment of [23]. However, the ripple in the experiment is much smaller than in the simulations [25, 26]. The suppression of the background ripple remains an open issue. In order to rule out the effect of higher dimensions, we have performed two-dimensional simulations in cylindrical coordinates which show the same behavior as the 1D simulations. We note that the ripple does grow significantly at late times in [23], which may be due to nonlinear backreaction from the self-amplifying Hawking radiation, as suggested in [24].

We thank Renaud Parentani for helpful comments. This work was supported by the Israel Science Foundation.